 \documentclass[pmlr,twocolumn]{jmlr} 

\usepackage{booktabs}
\usepackage[load-configurations=version-1]{siunitx} 

\theorembodyfont{\upshape}
\theoremheaderfont{\scshape}
\theorempostheader{:}
\theoremsep{\newline}

\jmlrvolume{}
\jmlryear{}
\jmlrsubmitted{}
\jmlrpublished{}
\jmlrworkshop{} 

\title{Improving Lesion Detection by exploring bias on \\ 
Skin Lesion dataset}

\author{%
\Name{Anusua Trivedi*} 
\\
\addr Microsoft, WA, USA
\AND
\Name{Sreya Muppalla*} 
\\
\addr Cupertino High School, CA, USA
\AND
\Name{Shreyaan Pathak*} 
\\
\addr Inglemoor High School, WA, USA
\AND
\Name{Azadeh Mobasher} 
\\
\Name{Pawel Janowski} 
\\
\Name{Rahul Dodhia} 
\\
\Name{Juan M. Lavista Ferres} 
\\
\addr Microsoft, WA, USA\\ \\
*These authors contributed equally as first authors
}
\begin{document}
\maketitle
\begin{abstract}
All datasets contain some biases, often unintentional, due to how they were acquired and annotated. These biases distort machine-learning models' performance, creating spurious correlations that the models can unfairly exploit, or, contrarily destroying clear correlations that the models could learn. With the popularity of deep learning models, automated skin lesion analysis is starting to play an essential role in the early detection of Melanoma. The ISIC Archive~\cite{codella2019skin} is one of the most used skin lesion sources to benchmark deep learning based tools. Bissoto et al. ~\cite{bissoto2019constructing} experimented with different bounding-box based masks and showed that deep learning models could classify skin lesion images without clinically meaningful information in the input data. Their findings seem confounding since the ablated regions (random rectangular boxes) are not significant. The shape of the lesion is a crucial factor in the clinical characterization of a skin lesion~\cite{silveria2009}. In that context, we performed a set of experiments that generate shape-preserving masks instead of rectangular bounding-box based masks. A deep learning model trained on these shape-preserving masked images does not outperform models trained on images without clinically meaningful information. That strongly suggests spurious correlations guiding the models. We propose use of general adversarial network (GAN) to mitigate the underlying bias.
\end{abstract}
\begin{keywords}
bias, health, skin, cancer
\end{keywords}
\section{Introduction}
Skin cancer is a growing problem globally, and it is the most common cause of cancer in the U.S ~\cite{guy2015vital}. The World Health Organization (WHO) predicts that a 10\% decrease in the depleting ozone layer will cause more than 300,000 additional cases of skin cancer ~\cite{who}. Like any other cancer, early lesion detection is key to improved clinical outcomes. However, lack of access to medical experts, especially in developing nations or areas of the world with inadequate screening facilities, implies higher mortality due to missed diagnosis or diagnosis once cancer has metastasized \cite{sandru2014survival}. Manually diagnosing is ineffective and expensive ~\cite{guy2015prevalence}.
Automated skin cancer lesion detection has potential benefits such as increasing efficiency, reproducibility, coverage of screening programs, reducing barriers to access, and improving patient outcomes by providing early detection and treatment~\cite{tripp2016state}. An algorithm to detect referable skin cancer is needed to maximize the clinical utility of automated lesion detection. 
Researchers have used machine learning for various classification tasks in recent years, including automated lesion detection for skin cancer. Identifying candidate regions in medical images is of the most significant importance since it provides intuitive illustrations for doctors and patients of inferred diagnosis. Recently, advances in Deep Learning have dramatically improved the performance of skin lesion detection. Most of these Deep Learning systems treat Convolutional Neural Network (CNN) as a black box, lacking comprehensive explanation. Diagnoses using deep neural nets with high precision and low recall on a mobile phone are feasible ~\cite{de2019development}. 
However, the state-of-the-art models are trained on mostly skin lesion datasets that belong to fair-skinned individuals rather than dark-skinned persons. An AI model trained on predominantly lighter skin background in the image may perform sub-optimally in darker skin backgrounds poorly represented in the training set. Even though the risk of developing skin cancer is relatively high among the light-skinned population, people with dark skin are also at risk. They are frequently diagnosed at later stages ~\cite{hu2006comparison}. Skin cancer accounts for 4 to 5\%, 2 to 4\%, 1 to 2\% of all cancers in Hispanics, Asians, and Blacks, respectively ~\cite{gloster2006skin}. Hence, deep learning frameworks validated for skin cancer diagnosis in fair-skinned people have a greater risk of misdiagnosing those with darker skin ~\cite{marcus2019rebooting}. In a recent study, Han et al. ~\cite{han2018classification} trained a deep learning algorithm on a training dataset consisting of skin lesions from Asians. They reported an accuracy of 81\% on the Asian testing set, whereas they reported accuracy of only 56\% on the Dermofit dataset, which consists of Caucasian people's skin lesions. Therefore, this drop-in accuracy signifies a lack of transferability of the learned features of deep learning algorithms across datasets that contain persons of a different race, ethnicity, or population. We argue that using the immediate area around the lesion can reduce the pressure on having a well-diversified skin color representation in the training data set. This approach's inspiration stems from the observation that physicians grading lesions are primarily driven by the morphology of the lesion and less so by the color of the skin surrounding the lesion \cite{silveria2009} \cite{amcansoc2019}. 
Gathering a fair amount of balanced benign and malignant skin lesion images for different skin profiles is a gigantic task. Bissoto et al. ~\cite{bissoto2019constructing} highlighted that the challenge is due to the vast visual variability of skin lesion and the subtlety of the cues that differentiate benign and malignant cases. They experimented with different bounding box based masks and showed that deep learning models could classify skin lesion images without clinically meaningful information. The shape of the lesion is a crucial factor in the clinical characterization of a skin lesion \cite{silveria2009}. In that context, we decided to extend the work by Bissoto et al. ~\cite{bissoto2019constructing} and experiment with shape-preserving masks. In this paper, we revisit a classical background segmentation approach using Otsu based adaptive thresholding and combine it with binary morphology operations to reduce noise while preserving the lesion shape in the image ~\cite{cheriet-1998, khan-2016}. 
\section{Dataset}
In this study we use an open-source dataset - HAM10000 ~\cite{tschandl2019ham10000}. HAM10000 dataset is commonly used as a benchmark database for academic machine learning purposes. We used the training set part of this
dataset consisting of 10015 dermatoscopic images with a size of 450x600. In this paper, we use a binary classification model, and we need to create a binary label - Benign vs. Malignant for the dataset. However, HAM10000 Metadata provides labels for different types of lesions classified by histopathology. For example, the BCC label stands for basal cell carcinoma, and MEL stands for melanoma, representing cancer. The label BKL stands for benign keratosis like lesions. This information is provided as a part of the Kaggle data set ~\cite{tschandl2019ham10000}. We used popular web search engines such as Bing and Google to map the labels provided to malignant or benign categories. This map is used to relabel the original labels into malignant and benign categories.  
\section{Image Processing Segmentation}
\begin{figure*}
\begin{center}
\graphicspath{ {./pics/} }
\includegraphics[scale=0.5]{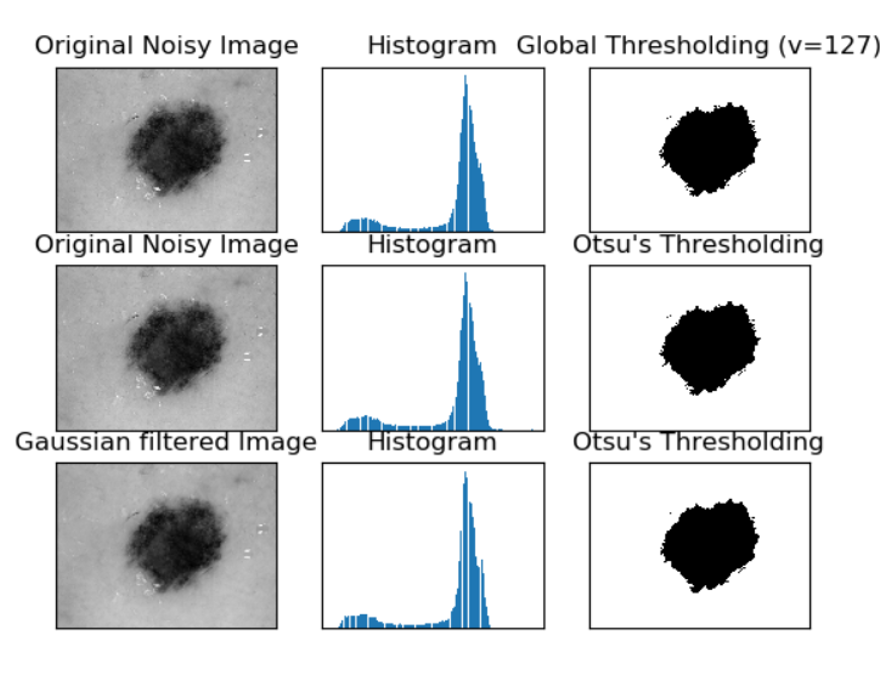}  
\caption{Different masking method. Top row: Global thresholding applied to original image, Middle row: Otsu thresholding applied to original image, Bottom row: Otsu thresholding applied to Gaussian filtered image.}
\label{fig:otsu}
\end{center}
\end{figure*}
\begin{figure*}
\begin{center}
\graphicspath{ {./pics/} }
\includegraphics[scale=0.5]{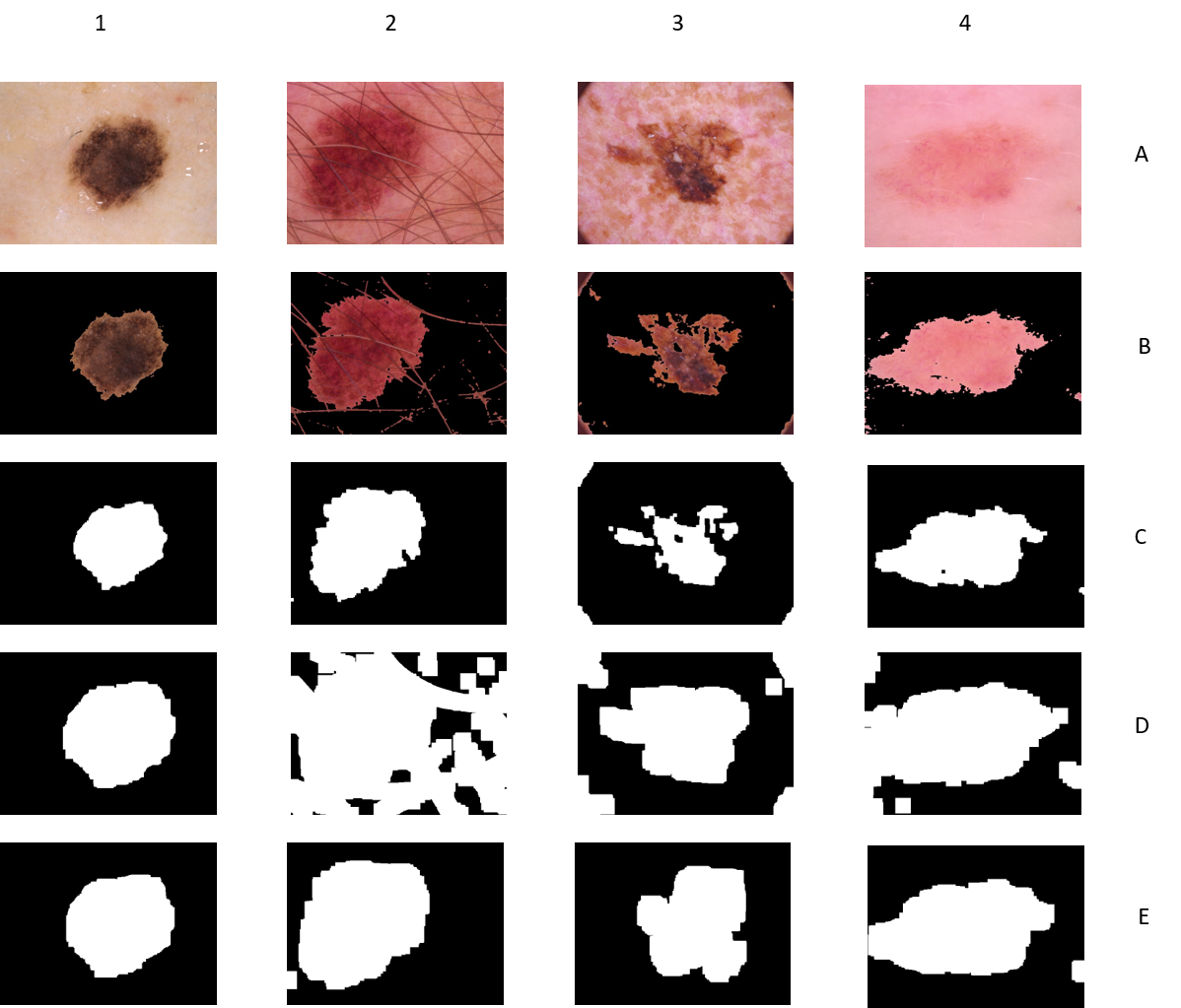}
\caption{Panel of 4 different skin lesion images along with their masked representations. Row A: Original images, Row B: Lesion with Otsu thresholded masks, Row C: Otsu masks masks cleaned with binary morphology erosion, Row D: Otsu masks dilated but not cleaned with binary morphology, Row E: Otsu masks cleaned with binary morphology opening and dilated to capture surrounding tissue}
\label{fig:skinpanel}
\end{center}
\end{figure*}
Segmentation and classification of skin lesions using deep learning techniques have been of key interest \cite{estevanature2017}. The basic segmentation is obtained by applying the following filters to the source image: 
\begin{itemize}
\item[(\textbf{1})] Gray-scale conversion of the color lesion image (Fig.~\ref{fig:otsu} column 1). We experimented with using the original image and also with Gaussian smoothing. 
\item[(\textbf{2})] Extract a binary mask using global thresholding and Otsu thresholding ~\cite{otsu94}. We found Otsu to be more robust to different image variations. It automatically can find the optimal partitioning point in the grayscale image into the foreground and background, as shown by the bi-modal histogram in the middle column of Figure ~\ref{fig:otsu}. 
\item[(\textbf{3})]We apply opening binary morphology operation ~\cite{harlickbinmorph87} to remove noisy regions (Fig.~\ref{fig:skinpanel}, row C). Figure.~\ref{fig:skinpanel}(C2) shows efficient hair removal while preserving the shape of the lesion. 
\item[(\textbf{4})] We apply dilation binary morphology operation ~\cite{harlickbinmorph87} to preserve the shape of the lesion and also capture the nearby region of the lesion similar to what a physician would do (Fig.~\ref{fig:skinpanel}, row D). However, as you can see, Fig.~\ref{fig:skinpanel}(D2-D4) dilation operation also increases the noise. Fig.~\ref{fig:skinpanel}(row E) shows how combining binary erosion and dilation can get accurate capturing of the skin lesion. In Figure~\ref{fig:skinpanel}(column 4), we can see even when the lesion is very faint, the segmentation mask captures the lesion area with reasonably good accuracy. 
\end{itemize}
\section{Lesion Classification using Preprocessed Images}
A key parameter when using binary morphology is the size and shape of the morphology kernel. Our goal was to generate different masks by varying the opening kernel's shapes and the dilation kernels. We varied the size of square kernel shapes (0, 5, 8, 10, 12, 15 pixels) and generated masks with different combinations (a total of 15 mask set) for the experiments. We trained a simple CNN model to prove the hypothesis using shape-preserving masks. 
We use an ImageNet~\cite{russakovsky2015imagenet} pre-trained VGG16~\cite{simonyan2014very} model and finetune it with our own data. VGG16 is a 16-layer CNN used by the Visual Geometry Group (VGG) at Oxford University. In 2014, the model won the 2nd place in the ILSVRC (ImageNet) competition, achieving a 7.5\% top 5 error rate on the validation set. After defining the fully connected layer architectures, we load the ImageNet pre-trained weight to the model. We then truncate the original softmax layer and replace it with our number of class labels. Then we finetune the model by minimizing the cross-entropy loss function using the stochastic gradient descent (SGD) algorithm. We keep the finetuned CNN very simple to run the evaluation quickly and directionally observe how changing the different mask sizes affect the accuracy of skin lesion characterization. 
\begin{table*}[!tb]
\begin{center}
\begin{tabular}{|c|c|c|}
\hline
\textbf{Experiment\#} & \textbf{\#Dilate\_\#Clean} & \textbf{TestAccuracy(in \%)} \\ \hline
\multicolumn{3}{l}{RGB Only} \\ \hline
1      & Baseline       & 82.2         \\ \hline
\multicolumn{3}{l|}{Mask Only}      \\ \hline 
2      & (10\_10)            & 80.88        \\ \hline
3      & (15\_15)            & 80.88        \\ \hline
4      & (10\_5)             & 80.88        \\ \hline
5      & (50\_80)            & 80.88        \\ \hline
\multicolumn{3}{l}{Mask with RGB}   \\ \hline
6      & (10\_10)            & 80.88        \\ \hline
7      & (15\_15)            & 81.7         \\ \hline
8      & (10\_5)             & 80.06        \\ \hline
9      & (50\_80)            & 80.88        \\ \hline
\end{tabular}
\caption {Summary of all the experiments using RGB image only (baseline), using only masks and using masks to cover lesion areas of the RGB images.}
\end{center}
\end{table*}
\begin{figure*}[t]
\begin{center}
\graphicspath{ {./pics//} }
\includegraphics[scale=0.5]{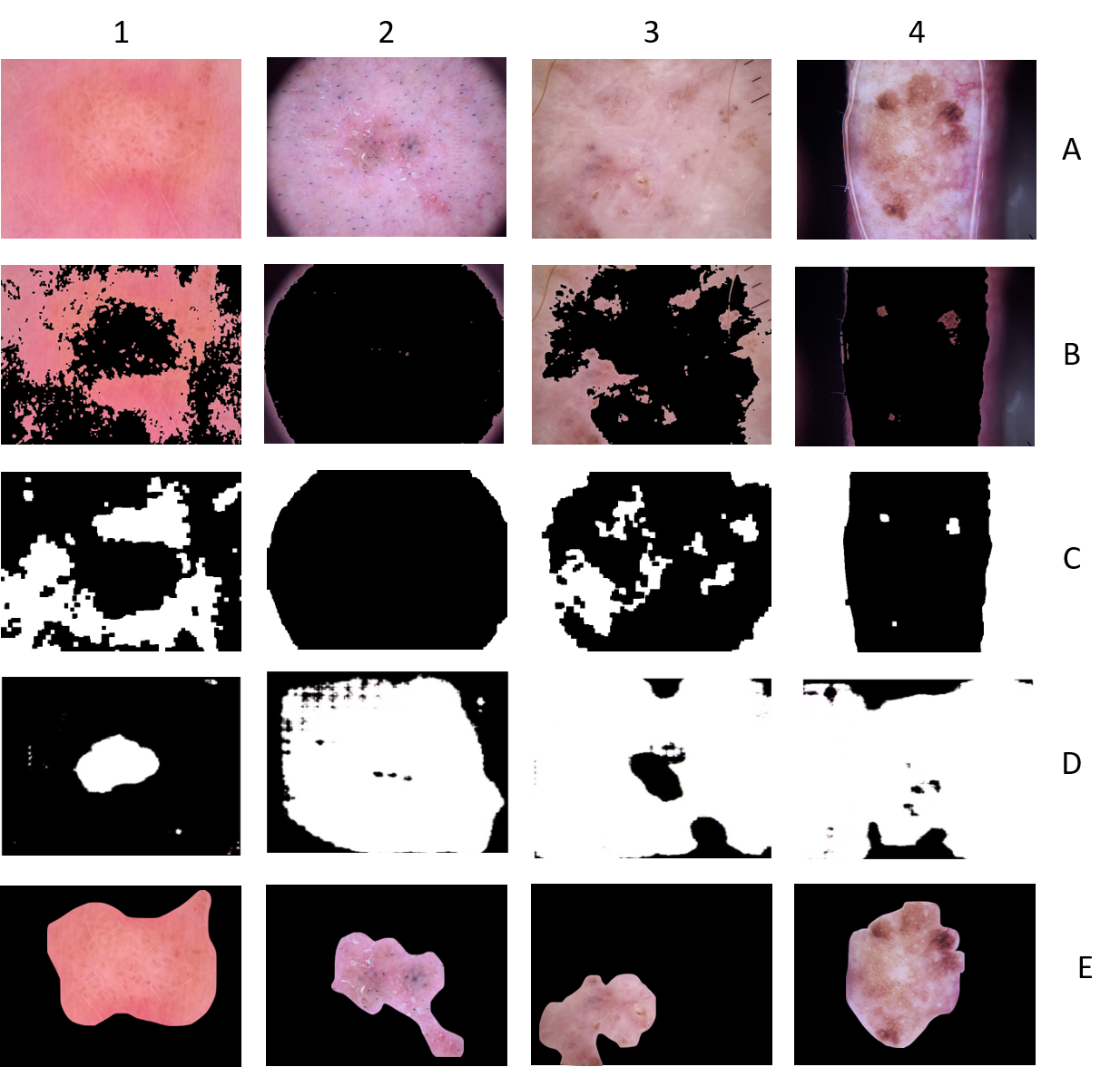}
\caption {Panel of malignant and benign lesions processed with otsu thresholding which failed to generate accurate masks. 1 and 2: Benign, 3 and 4: Malignant, A: Original Image, B: Image with Mask + RGB, C: Mask only with cleaning from 10 pixel kernel D: Mask generated through Pix2PixGAN E: Ground truth images outlined by a radiologist}
\label{fig:nosegexample}
\end{center}
\end{figure*}
In Table 1, we explore different mask-only and RGB with mask options with CNN. In experiment 1, we ran a baseline test to see how well the RGB does. Although the numbers might seem promising, the model does not take into account different skin tones. In experiments 2, 3, and 4, we started experimenting with varying commands of morphology. After trying different amounts of the mask to maximize the accuracy, we noticed that the accuracy numbers were coming out to be very similar, irrespective of the various masks. We also tried with extreme numbers such as experiment 5, but identical accuracy numbers showed up. This result confirmed the results shown by Bissoto et al. ~\cite{bissoto2019constructing}.  Next, we experimented with shape-preserving masks. In experiments 6,7 and 8, we tried using the mask on top of the RGB images. However, the accuracy numbers came in a similar range of 80/82 percent accuracy as with mask-only experiments in 2, 3, and 4. We also tried extreme masking with RBG, as shown in experiment 9, but similar accuracy numbers showed up. In conclusion, regardless of the amount of dilation, the classifier's accuracy does not change, remaining between (81\%-82.7\%). 
Additionally, in Figure~\ref{fig:nosegexample}, we can see the segmentation algorithm fails in images that have low contrast or very fuzzy boundaries. Fig~\ref{fig:nosegexample}A show very poor image contrast. The mask captures the periphery of the lesion, but the low contrast and low-intensity interior are missed. This is clearly illustrated in  Fig~\ref{fig:nosegexample} row B1-3 and C1-3. Fig~\ref{fig:nosegexample}A2 is so subtle that the mask completely fails. Note row E corresponds to outlines drawn by a certified Radiologist. For these four images, the radiologist remarked that they were challenging because of poor contrast and, in some cases, skin damaged due to scratching or inflammatory effects from localized irritations. The radiologist also consulted another colleague with dermatology experience. They concluded that for these types of samples, the inter-expert disagreement would be very high.
\section{Generative Model based Segmentation} 
As we show with the demonstration mentioned above, skin cancer lesion classification fails with simple segmentation masks, and we want to evaluate more sophisticated deep generative models like Pix2PixGAN~\cite{isola2017image}. In recent years, deep learning is shown to be successful in visual perception and the semantic segmentation of images. General Adversarial Networks or GANs can be briefly divided into two parts - encoder and decoder. The encoder is a convolution neural network that extracts features from the input image, such as the skin image's features. The decoder will up-sample the extracted features to the resulting image that we desired as the skin lesion segmentation in our case. Researchers have already used variations of U-NET~\cite{ronneberger2015u}, Residual U-NET~\cite{zhang2018road}, and R2U-Net~\cite{alom2018recurrent} on ISIC data set as shown in ~\cite{skin}. However, regardless of race/background, we want anyone to have immediate and less costly access to find out if they had skin cancer. Bissoto et al. ~\cite{bissoto2018skin} have shown the use of Pix2PixHD GAN for skin images. However, it requires the use of high quality altered images, which is not a guarantee for people in developing nations without access to technology equipped with better quality cameras. This leads us to the idea of improving lesion detection without bias by using Pix2PixGAN. We trained a Pix2PixGAN to make the results more precise and universal.  
\begin{figure*}
\begin{center}
\graphicspath{ {./pics//} }
\includegraphics[scale=0.7]{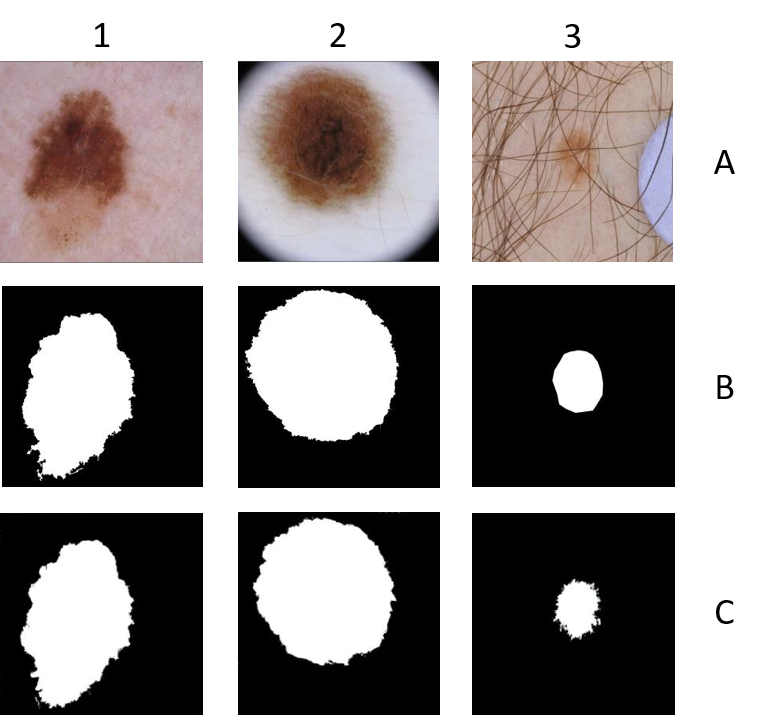}
\caption {Panel of segmented skin lesion images generated by Pix2PixGAN compared to its ground truth.  Row A: Original RGB image, Row B: Ground Truth, Row C: Predicted image with Pix2PixGAN}
\label{fig:nosample}
\end{center}
\end{figure*}
\begin{figure*}
\begin{center}
\graphicspath{ {./pics//} }
\includegraphics[width=1\linewidth]{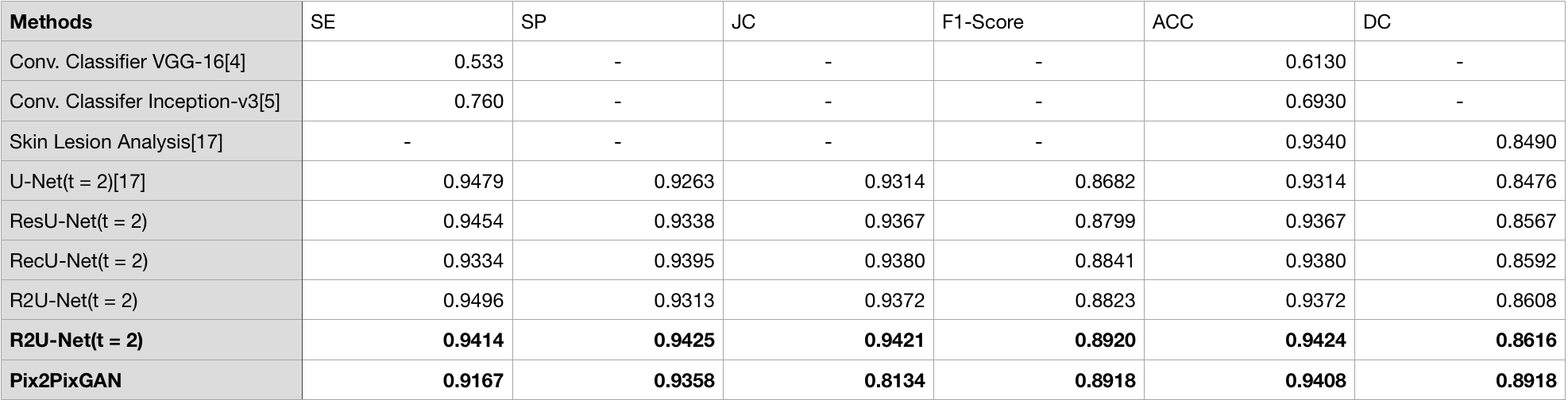}
\caption{Experimental Results Of Proposed Approaches For Skin Cancer Lesion Segmentation And Comparison Against Other Existing Approaches}
~\cite{burdick2018rethinking}
~\cite{codella2018skin}
~\cite{li2018skin}
\label{fig:table2}
\end{center}
\end{figure*}
In Fig~\ref{fig:nosample}, it shows the results of our trained Pix2PixGAN model on the ISIC dataset. We chose different types of images of skin lesions to point out that Pix2PixGAN was able to segment the correct part of the lesion regardless of the background. In column 1, the model was able to correctly detect the lesion, matching exactly to the ground truth of the corresponding image. In column 2, Fig~\ref{fig:nosample}, there is a bit of a background for the skin lesion. However, the GAN was still able to predict the segmented image flawlessly. The skin lesion in column 3 Fig~\ref{fig:nosample} proves how accurate our model is because of the noises in the image, such as hair and random patch in the corner. We cannot always guarantee that users will have images with no distraction, and Pix2PixGAN proves to be still successful under those situations. Figure 4 Fig~\ref{fig:nosample} also supports the strong evaluation metrics provided in table 2, as evidenced by the model's accuracy. 
In Fig~\ref{fig:nosegexample}, we compare the mask corresponding to the input images compared to the Pix2PixGAN(using the Kaggle dataset). In all the columns(except column 3), Pix2PixGAN was able to detect the skin lesion at a much higher accuracy than the mask. The mask(Row C) detected the background in the images, whereas the GAN was able to detect parts of the skin lesion. In columns 1 and 2, Fig~\ref{fig:nosample}, Pix2PixGAN  was able to detect the lesion pretty accurately. However, in column 3 Fig~\ref{fig:nosample}, the GAN was unable to detect the skin lesions due to them being scattered on the image, and the model was never trained in such cases. In the last column, Fig~\ref{fig:nosample}, again, the GAN was met with an unexpected image and was unable to detect the skin lesion. However, the model produced a better image than the mask(only to some extent). The Pix2PixGAN model could not detect as accurately because it was met with an unexpected input that it was not trained on. We shared these images with two expert doctors with experience in radiology and dermatology. After manually outlining the images, they confirmed that these images were challenging to outline even humans. This indicates our approach is robustly tested against diverse test cases. Fig~\ref{fig:nosegexample}D(2-4) shows the GAN labeled regions were more expansive. This gives us an additional signal in the application to flag such images for manual intervention. For Fig~\ref{fig:nosegexample}D(1), the GAN outlined the lesion area while the physician included additional areas around the lesion. Such subjective variations in interpretation make skin-lesion detection challenging even for well-trained specialists. We plan to incorporate this additional clinical insight into future GAN refinements.   
As seen in Fig~\ref{fig:table2}, different types of metrics were used to provide quantitative results with the Pix2PixGAN model: Accuracy(ACC), Sensitivity(SE), Specificity(SP), F-1 score, and Dice Coefficient(DC). Compared to the R2-Unet model, Pix2PixGAN had approximately the same accuracy(0.94). Accuracy is calculated using this formula:(TP + TN)/(TP + TN + FP + FN). TP is true positive; TN is true negative; FP is false positive; FN is false negative. Moreover, the high accuracy of the Pix2PixGAN model proves it can perform the segmentation task more effectively than the U-Net and ResUnet models, which had lower accuracies. Pix2PixGAN's accuracy portrays its accurate results in end-to-end segmentation tasks.  The sensitivity of the Pix2PixGAN model was 0.91, and the specificity of the model was roughly 0.93. The formula used to calculate sensitivity is \textbf{TP/(TP + FN)}. The formula used to calculate specificity is \textbf{TN/(TN + FP)}. Pix2PixGAN's F1-score proves the model as successful as R2-Unet, as both have similar scores(0.89). The F1-score takes the average of both precision(for Pix2PixGAN, it was 0.89) and recall. The Dice Coefficient(DC) of Pix2PixGAN is 0.89, which is higher than R2-Unet's(0.86). The DC score stands for Dice Coefficient, which is the size of the overlap of the two segmentations divided by the two objects' total size. The DC score is often used as a reproducibility validation metric, denoting that Pix2PixGAN would produce these segmented images with the same accuracy multiple times, whereas R2-Unet would have a lower chance of doing that. This is especially important if multiple people will be using this app, as Pix2PixGAN will be more reliable for producing segmented images.
\section{Conclusion}
Bissoto et al. ~\cite{bissoto2019constructing} experimented with different bounding-box based masks and showed that deep learning models could classify skin lesion images without clinically meaningful information. Our result extends their work and shows that even shape-preserving masks fail to improve classification accuracy over deep learning models that classify skin lesion masked images without clinically meaningful information. We conclude that the simple CNN type classifiers are by themselves inadequate in modeling skin cancer lesions. However, generating robust segmentation is critical to remove bias in skin cancer data. In our data set, we also found that a handful of images showed partial or no segmentation at all due in part to image quality, intensity variation, lens artifacts \textit{etc}. Although the number of failures is small but due to the nature of the problems with lives at stake, we need a better segmentation algorithm that is more adaptable to subtle variations in skin lesion color/shapes and generalizes well in the face of wider variations within the population. The combination of segmented skin lesions along with a more extensive data set (several hundreds of thousands) ~\cite{estevanature2017}, is likely to be a winning combination. 
\acks{We sincerely thank Dr. Sayan Pathak, Affl Prof of CSE at Indian Institute of Technology and former faculty at Dept. of Bioengineering at the University of Washington, Seattle, WA, for his invaluable guidance and insights. We also thank Dr. Manjari Gajra of School of Medical Sciences, University of Manchester and Dr. Ryan Pathak, Consultant Radiologist at Salford Royal Hospital for their manual annotation of the lesions and evaluation of the predicted annotations.}
\bibliography{jmlr-sample}
\end{document}